\documentclass[%
 reprint,
 amsmath,amssymb,
 aps,
pra,
]{revtex4-2}

\usepackage{graphicx}
\usepackage{dcolumn}
\usepackage{bm}

\begin{document}

\preprint{APS/123-QED}

\title{Constraints on Meso-Scale Structure in Complex Networks}

\author{Rudy Arthur}
 \email{R.Arthur@exeter.ac.uk}
\affiliation{%
 University of Exeter, Department of Computer Science,\\
Stocker Rd, Exeter EX4 4PY
}%

\date{\today}

\begin{abstract}
A key topic in network science is the detection of intermediate or meso-scale structures. Community, core-periphery, disassortative and other partitions allow us to understand the organisation and function of large networks. In this work we study under what conditions certain common meso-scale structures are detectable using the idea of block modularity. We find that the configuration model imposes strong restrictions on core-periphery and related structures in directed networks. We derive inequalities expressing when such structures can be detected under the configuration model. Nestedness is closely related to core-periphery and is similarly restricted to only be detectable under certain conditions. We show that these conditions are a generalisation of the resolution limit to structures other than assortative communities. We show how block modularity is related to the degree corrected Stochastic Block Model and that optimisation of one can be made equivalent to the other in general. Finally, we discuss these issues in inferential versus descriptive approaches to meso-scale structure detection.
\end{abstract}

\maketitle

\section{Introduction}\label{sec:introduction}

An important topic in network science is the identification of intermediate or meso-scale structure by partitioning the nodes of the network into non-overlapping sets \cite{barabasi2013network,fortunato202220}. The primary structure of interest has historically been assortative communities \cite{radicchi2004defining,newman2004detecting,fortunato2010community,fortunato2016community}. These are tightly connected groups of nodes which split the network into a number of weakly connected modules. Another meso-scale structures of interest is core-periphery \cite{borgatti2000models,csermely2013structure,rombach2014core,zhang2015identification,kojaku2017finding,yanchenko2023core}, consisting of a core set and a periphery set, where the core is densely connected internally and to the periphery, while periphery nodes are predominantly connected to core nodes. There are also disassortative structures which have received somewhat less attention \cite{holme2003network,estrada2005spectral,newman2006finding,barber2007modularity,chen2014anti,arthur2020modularity}, but represent the other side of community structure, dense connections between groups and sparse connections within them. 

Recent work \cite{kojaku2018core, li2021anomaly,liu2023nonassortative} has sought to classify meso-scale structures which partition the nodes of the network into non-overlapping sets. This means structures like k-cores \cite{alvarez2005large} and rich-clubs \cite{zhou2004rich} are not considered. However, when we have a hard partition the network's adjacency matrix has an approximate block structure. For small numbers of subsets (2 in \cite{li2021anomaly} and 2,3,4 in \cite{kojaku2018core}) it is feasible to enumerate all the possible blocks.  \cite{li2021anomaly,liu2023nonassortative} identify meso-scale structure based on the ratio of existing to possible links within and between communities. \cite{kojaku2018core} bases meso-scale structure detection on the excess or deficit of links within or between blocks compared to a null model. We will follow the latter approach in this work, which allows us to study the assumptions required to determine if a network can have a particular structure.

Building on \cite{kojaku2018core}, in \cite{arthur2023discovering} I defined a generalised notion of modularity called \emph{block modularity}. This extends the original modularity definition \cite{newman2006modularity} in a way that allows the evaluation of partition quality for a variety of meso-scale structures other than assortative communities. Block modularity captures the intuitively reasonable notion that, for us to declare that a meso-scale structure is present there should be more links within or between the corresponding blocks than a random network would have. A key point is that the `random network' is one selected from a particular null model. While numerous choices for a null model are possible: Erd\H{o}s-R\`enyi (ER), Stochastic Block Model (SBM) \cite{karrer2011stochastic} or others \cite{chung2002average,expert2011uncovering}, generally the configuration model \cite{newman2001random} is used. 

A random network drawn from the configuration null is one with the same degree sequence as the observed network. Degree distribution, along with correlations and anti-correlations among node degrees have long been recognised as key characteristics that explain many of the properties of complex networks \cite{albert2002statistical,park2003origin, barabasi2013network}. Significant differences between the observed network and its configuration model implies that there is structure in the network which cannot be explained by the node degrees alone. This requirement of structure beyond that implied by the degree sequence turns out to be quite rigid, and imposes hard constraints on the types of meso-scale structures which are possible. In particular \cite{kojaku2018core} found that a partition of a network into one core and one periphery set is not possible under the configuration null.

This work builds on and extends this result. First, in Section \ref{sec:blockmod} we introduce block modularity and extend it to work for directed networks. Section \ref{sec:allowed} enumerates all possible $2 \times 2$ block patterns for directed and undirected networks. We then use block modularity to rederive the constraints found by \cite{kojaku2018core} in a somewhat simpler way, as well as extending those results to the directed case.  In Section \ref{sec:constraint} we derive further restrictions on Core-Periphery structure (as well as a directed network structure we refer to as Community-Hierarchy) when these structures are part of larger networks. Section \ref{sec:nestedness} discusses nested networks \cite{konig2014nestedness} under the configuration model. In Section \ref{sec:resolution} we show how these restrictions are related to and generalise the resolution limit \cite{fortunato2007resolution}.
Section \ref{sec:SBM} considers another popular null model, the degree corrected Stochastic Block Model (dc-SBM) \cite{karrer2011stochastic}. We show how this is related to Block Modularity and how structure detection with the dc-SBM differs from modularity based approaches.

\section{Block Modularity}\label{sec:blockmod}
Modularity was introduced by Newman  \cite{newman2006modularity} to measure the quality of a partition of a network into communities
\begin{equation}\label{eqn:Qnewman}
    Q_{Newman} = \frac{1}{2E}\sum_{ij} \left(A_{ij} - P_{ij} \right) \delta( c(i), c(j) )
\end{equation}
The sums are over all $N$ nodes of the network, $A$ is the adjacency matrix, $2E = \sum_{ij} A_{ij}$ is twice the number of edges, $c(i)$ is a function which returns the community label of the node $i$ and $P_{ij}$ is the expected number of links between $i$ and $j$ in the null model. With the configuration model as the null
\[
P_{ij} = \frac{k_i k_j}{2E}
\]
where $k_i = \sum_j A_{ij}$ is the degree of $i$. An Erd\H{o}s-R\`enyi (ER) null model, which assumes a constant probability for every edge to exist (including self edges) has
\[
P_{ij} = p = \frac{2E}{N^2}
\]
Another common null is the scaled configuration model \cite{reichardt2006statistical}
\[
P_{ij} = \gamma \frac{k_i k_j}{2E}
\]
where $\gamma$ is an arbitrary parameter. We will primarily deal with the configuration model and discuss the ER and scaled configuration model in Appendix \ref{sec:appendix}. The stochastic block model (SBM) \cite{lee2019review} is also often used to investigate meso-scale structure, though in general not together with modularity. We treat the SBM in Section \ref{sec:SBM}.

Modularity can be rewritten in terms of the adjacency matrix of the network induced by the partition. Define
\[
S_{ab} = \sum_{i \in a, j \in b} A_{ij}
\]
as the number of edges going between $a$ and $b$, or twice the number of edges inside $a$ if $a = b$. Similarly for the null define,
\[
S_{ab}^P = \sum_{i \in a, j \in b} P_{ij}.
\]
In these terms, Equation \ref{eqn:Qnewman} becomes
\begin{equation}
    Q_{Newman} = \frac{1}{2E}\sum_{a} \left( S_{aa} - S^P_{aa} \right)
\end{equation}
where the sum is over all $K$ sets in the partition. Written like this, $Q_{Newman}$ is the sum over groups of the excess or deficit of edges in the observed network compared to the number predicted by the null model. For the configuration null
\[
    Q_{Newman} = \frac{1}{2E}\sum_{a}^K \left( S_{aa} - \frac{T_a^2}{2E} \right)
\]
with 
\[
T_a = \sum_{i \in a} k_i
\]

\cite{arthur2023discovering} introduced a natural generalisation of this function (see also \cite{kojaku2017finding}), with
\[
    Q_{ab} = S_{ab} - S^P_{ab}
\]
the \emph{block modularity} is defined as
\begin{equation}
    Q(B) = \frac{1}{2E}\sum_{ab}^K Q_{ab} B_{ab}
\end{equation}
with the \emph{block matrix} $B$ having entries equal to $\pm 1$. For the configuration model, 
\[
S^P_{ab} = \frac{T_a T_b}{2E}
\]
The target meso-scale structure is encoded in the block matrix $B$. For example, if the matrix $B$ has 1s on on the diagonal and -1s elsewhere $Q(B) = 2 Q_{Newman}$ and $Q(B)$ evaluates how well a partition groups the nodes into assortative communities. However we are not limited to assortative communities. For example, if 
\[
B = \begin{pmatrix}
    1 & 1 & -1 & -1 \\
    1 & -1 & -1 & -1 \\
    -1 & -1 & 1 & 1 \\
    -1 & -1 & 1 & -1 
\end{pmatrix}
\]
then $Q(B)$ evaluates the quality of a partition of the nodes into a pair of core-peripheries. \cite{arthur2023discovering} explores simultaneously optimising both the partition $c(i)$ and the block pattern $B$, so that the meso-scale structure is discovered from the data and in Section \ref{sec:SBM} we will discuss an algorithm for determining $B$. However, for a given target structure we don't need to learn $B$ from the data, we can simply choose $B$ to probe the network for the structure of interest.

\subsection{Directed networks}

Block modularity can be extended straightforwardly to directed networks in much the same way as modularity is extended in \cite{nicosia2009extending}. Let $A$ be the adjacency matrix of a directed graph. Recall, in the directed case $A_{ij} \neq A_{ji}$. With
\begin{align*}
k^{(in)}_j = \sum_i A_{ij},\quad
k^{(out)}_i = \sum_j A_{ij},\quad
E = \sum_{ij} A_{ij}
\end{align*}

Define
\begin{equation}
    Q_{\leftrightarrow}(B) = \frac{1}{E}\sum_{ab}^K Q_{ab} B_{ab}
\end{equation}
with $Q_{ab}$ and $B_{ab}$ as before, noting that in the directed case these need not be symmetric. Using the directed configuration model as the null
\begin{align*}
S^P_{ab} = \frac{T_a^{(out)} T_b^{(in)}}{E},\quad
T_a^{(out)} = \sum_{i \in a} k^{(out)}_i,\quad
T_b^{(in)} = \sum_{j \in b} k^{(in)}_j
\end{align*}

\section{Allowed Block Patterns}\label{sec:allowed}

\begin{figure*}
    \centering
    \includegraphics[width=0.8\textwidth]{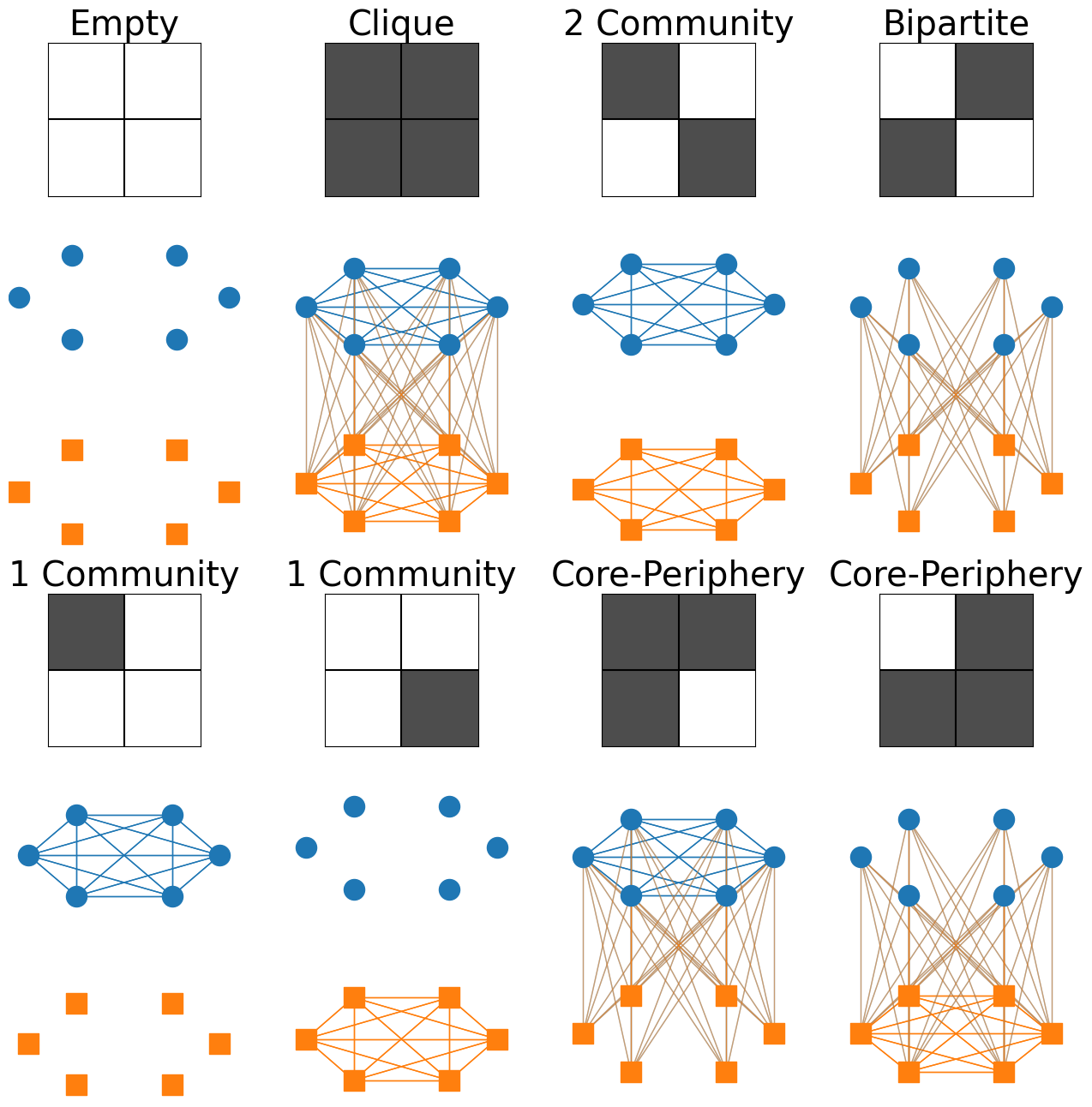}
    \caption{All 2x2 block patterns for undirected networks and an example network realising that block pattern.}
    \label{fig:22undirected}
\end{figure*}

\begin{figure*}
    \centering
    \includegraphics[width=0.6\textwidth]{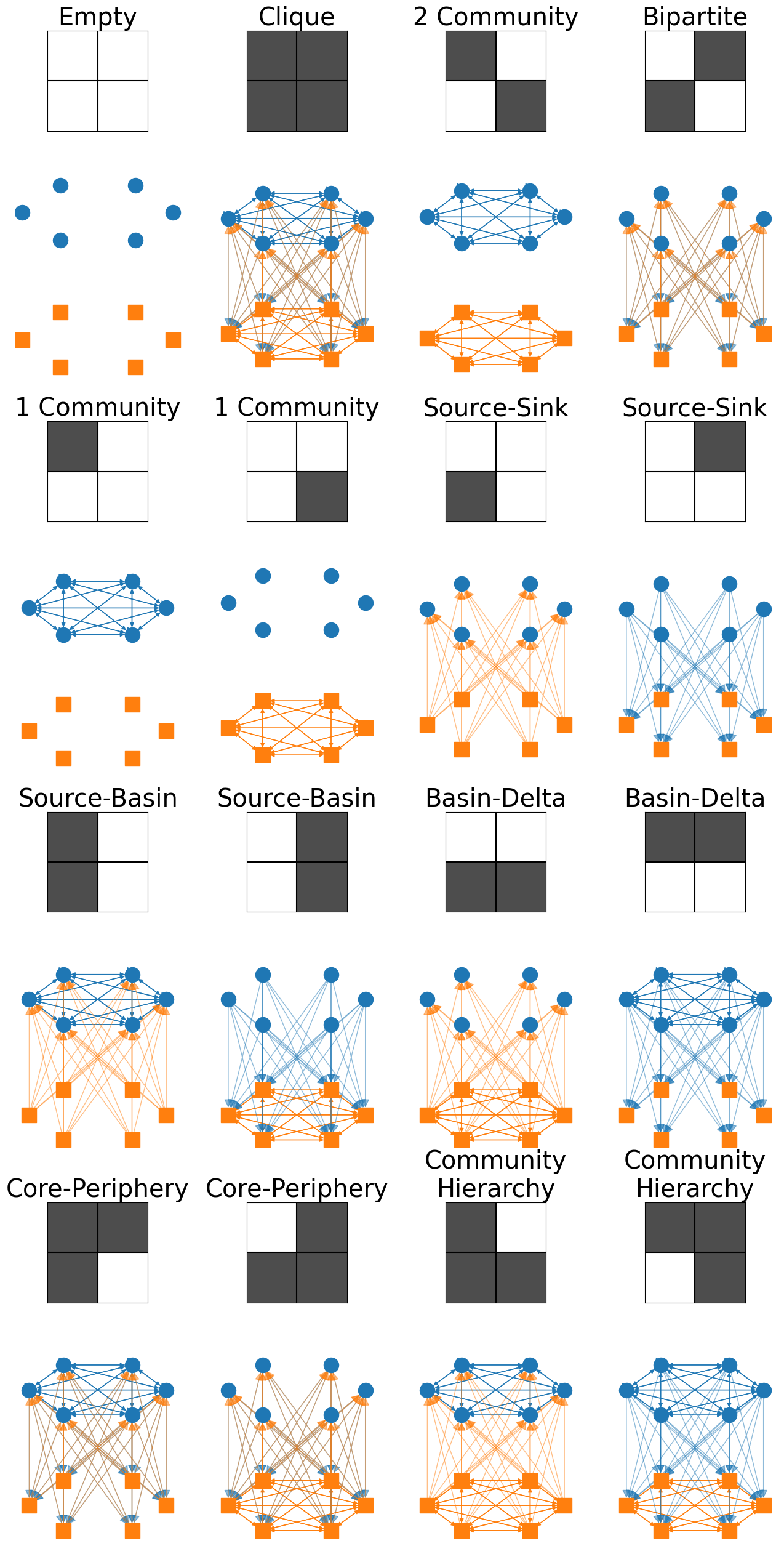}
    \caption{All 2x2 block patterns for directed networks and an example network realising that block pattern.}
    \label{fig:22directed}
\end{figure*}

All possible $2 \times 2$ block matrices $B$ with entries $\pm 1$ are shown in Figure \ref{fig:22undirected} for the undirected and Figure \ref{fig:22directed} for the directed case, where a black square is $+1$ and a white square $-1$, corresponding to an excess or deficit of edges respectively. The example networks show the limit where the $+1$ blocks are fully connected and the $-1$ blocks have no edges. \cite{liu2023nonassortative} groups the bottom 8 directed networks into two classes, source-basin and core-periphery. Here we split those groups into 
\begin{itemize}
    \item \textbf{Source-Basin}: one-way flow from a loosely connected group to a densely connected group.
    \item \textbf{Basin-Delta}: one-way flow from a densely connected group to a loosely connected group.
    \item \textbf{Core-Periphery}: two-way flow between a densely connected group and a loosely connected group.
    \item \textbf{Community-Hierarchy}: one-way flow between densely connected groups.
\end{itemize}

\subsection{ Detectable $2 \times 2$ Patterns}

In the undirected case with the configuration model null
\begin{align}\label{eqn:KMrule}
    \sum_{b} Q_{ab} &= \sum_{b} S_{ab} - \frac{T_a }{2E} \sum_b T_b \\ \nonumber
    &= \sum_{i \in a, j } A_{ij} - \frac{T_a }{2E} \sum_{i \in a, j} k_j \\ \nonumber
    &= \sum_{i \in a }k_i - T_a = 0
\end{align}
This implies the row and column sums of the $Q_{ab}$ matrix can't simultaneously all be positive or all negative, which is the result found in \cite{kojaku2018core}. This means that we can't have a solid black or solid white, row or column in the matrices of Figure \ref{fig:22undirected}, otherwise $Q(B)$ is identically zero. In other words, it is impossible to have an excess (or deficit) of connections between one group and all the other groups under the configuration model null. I will refer to this as the KM rule. 

An almost exactly analogous calculation applies in the directed case
\begin{align*}    
\sum_{b} Q_{ab} &= \sum_{b} S_{ab} - \frac{T_a^{(out)} }{E} \sum_b T_b^{(in)}\\
       &= \sum_{i \in a, j } A_{ij} - \frac{T_a^{(out)} }{E} \sum_{i \in a, j} k^{(in)}_j \\
&= \sum_{i \in a } k_i^{(out)} - T_a^{(out)} = 0       
\end{align*}
and similarly for the column sums.  Thus in both the directed and undirected cases, under the configuration model the only detectable $2 \times 2$ block patterns are 2-Community and Bipartite (called disassortative in \cite{liu2023nonassortative}) networks
\[
\begin{pmatrix}
    1 & -1 \\
    -1 & 1
\end{pmatrix} \quad
\begin{pmatrix}
    -1 & 1 \\
    1 & -1
\end{pmatrix}
\]
In particular a $2 \times 2$ Core-Periphery is excluded.

With more than two blocks, other patterns can be observed, provided that the KM rules are satisfied. \cite{kojaku2018core} provides a list of all allowed $3 \times 3$ and $4 \times 4$ patterns for the undirected case. To see that there must be more than just these row and column sum constraints, consider a pattern like
\[
\begin{pmatrix}
    1 & 1 & -1\\
    1 & -1 & -1\\
    -1 & -1 & 1
\end{pmatrix}
\]
which represents a CP+1 community structure. This satisfies all the KM rules, however these rules say nothing about the sizes of the communities. Thus, an arbitrarily large $2 \times 2$ CP structure plus one isolated node would be allowed. We investigate the resolution of this paradox in the next section.

\section{Constraint Equations}\label{sec:constraint}

We will mostly consider the configuration model null, since this is the most common, and interesting, case. Other cases are considered in Appendix \ref{sec:appendix}. For a group, or pair of groups, to make a positive contribution to the modularity requires $Q_{ab} B_{ab} > 0$. Again, this is just the demand that in order to say a structure is really present, there should be an excess or deficit of connections within or between groups compared to a random network. The structure we are aiming to detect is encoded by $B$. So for example, assuming groups $c$ and $p$ formed a CP pair (as part of a larger network in order to satisfy the KM rules) we would require
\[
Q_{cc} > 0, Q_{cp} > 0, Q_{pp} < 0
\]
That is, an excess of connections within the core, from core to periphery and a deficit of connections within the periphery.

Working with the configuration null, we derive general conditions first, then apply them to specific structures in the relevant sections. We first write the $Q_{aa}$ and $Q_{ab}$ matrix elements in terms of $S_{aa}, S_{ab}$ and $S_{bb}$.  The degree sum can be split as follows,
\begin{align}
T_a &= \sum_{i \in c, j} A_{ij} = \sum_{i \in a, j \in a} A_{ij} + \sum_{i \in a, j \in b} A_{ij} + \sum_{i \in a, j \not\in a,b } A_{ij} \\ \nonumber
&= S_{aa} + S_{ab} + S_{a*}
\end{align}
where $S_{a*}$ is the number of links from group $a$ to all other groups apart from $a$ and $b$. We also write
\begin{equation}
2E = S_{aa} + S_{bb} + 2(S_{ab}  + S_{a*} + S_{b*}) + S_{**}
\end{equation}
Where $S_{**}$ is twice the number of links in and between all groups apart from $a$ and $b$. This gives 
\begin{align*}
    Q_{ab} = S_{ab} - \frac{(S_{aa} + S_{ab} + S_{a*})(S_{bb} + S_{ab} + S_{b*})}{ S_{aa} + S_{bb} + 2(S_{ab}  + S_{a*} + S_{b*}) + S_{**}}
\end{align*}

Some simple algebra the shows that the conditions for the matrix elements to be greater than zero can be written as
\begin{align}\label{eqn:fullQconditions}
Q_{aa} > 0 &\equiv S_{aa}S_{**} > (S_{ab} + S_{a*})^2 - S_{aa}(S_{bb} + 2S_{b*})\\
Q_{ab} > 0 &\equiv S_{ab} S_{**} > 
(S_{aa}+S_{a*})(S_{bb} + S_{b*}) \\ \nonumber
&\qquad\quad\qquad-S_{ab} ( S_{ab}+S_{a*}+S_{b*}) 
\end{align}
Consider the ideal case where $a$ and $b$ may be connected to each other but share no links with the rest of the network. Any outside connections will increase the null model term, so only makes it more difficult for the pair $a,b$ to make a positive contribution to $Q(B)$. In this case $S_{a*} = S_{b*} = 0$ and the above inequalities simplify to
\begin{align}\label{eqn:Qconditions}
Q_{aa} > 0 &\equiv S_{aa}S_{**} > S_{ab}^2 - S_{aa} S_{bb}\\
Q_{ab} > 0 &\equiv S_{ab} S_{**} > 
S_{aa}S_{bb} - S_{ab}^2 
\end{align}

\subsection{Core-Periphery Constraints}
Consider a block matrix which has CP structure between groups labelled $c$ and $p$, together with some other blocks that allow the KM rules to be satisfied. The block matrix to detect this looks like
\[
B = \begin{pmatrix}
\vdots & \vdots & \vdots & \vdots & \vdots \\
\ldots & 1 & \ldots & 1 \ldots \\
\vdots & \vdots & \vdots & \vdots & \vdots \\
\ldots & 1 & \ldots & -1 \ldots \\
\vdots & \vdots & \vdots & \vdots & \vdots \\
\end{pmatrix}
\]
The groups $c$ and $p$ make a positive contribution to $Q(B)$ if $Q_{cc}, Q_{cp} \geq 0$ and $Q_{pp} \leq 0$ (also assuming they are not all zero). In the ideal case there are no internal links in the $p$ group, so $S_{pp} = 0$. Assume also that there are no links between $c,p$ and the rest of the network. In this case $Q_{pp} < 0$ and $Q_{cp} > 0$ always. However we only have $Q_{cc} > 0$ if
\begin{equation}\label{eqn:cpconditions}
    S_{cc}S_{**} > S_{cp}^2 
\end{equation}
otherwise $Q_{cc} < 0$. If $Q_{cc} < 0$ then we have a deficit of edges in the core compared to what would be expected under the null. In this case a block matrix with $B_{cc} = -1$ would give a larger value of $Q(B)$, i.e. a bipartite (disassortative) relationship between $c$ and $p$ is less expected under the configuration null. A core-periphery is possible under the configuration model when
\begin{itemize}
    \item The number of edges in the core, $S_{cc}$, is large.
    \item The number of edges between core and periphery, $S_{cp}$, is small.
    \item The core-periphery system is part of a much larger network, $S_{**} \gg 0$.
\end{itemize}

\subsection{Directed Network Constraints}
The directed case can be analysed similarly and the equivalent of Equations \ref{eqn:Qconditions} and 11 for directed networks is
\begin{align}
Q_{aa} > 0 &\equiv S_{aa}S_{**} > S_{ab}S_{ba} - S_{aa} S_{bb}\\
Q_{ab} > 0 &\equiv S_{ab} S_{**} >
S_{aa}S_{bb} - S_{ab}S_{ba}
\end{align}
where it should be recalled that $S_{ab} \neq S_{ba}$ in general. CP structure is subject to the same constraints as in the undirected case. For a directed core-periphery to be detectable under the configuration model, we need
\begin{equation}\label{eqn:cpdirconditions}
    S_{cc}S_{**} > S_{cp}S_{pc}
\end{equation}

The community-hierarchy (CH) structure is another interesting case. The corresponding block matrix is
\[
B = \begin{pmatrix}
\vdots & \vdots & \vdots & \vdots & \vdots \\
\ldots & 1 & \ldots & 1 \ldots \\
\vdots & \vdots & \vdots & \vdots & \vdots \\
\ldots & -1 & \ldots & 1 \ldots \\
\vdots & \vdots & \vdots & \vdots & \vdots \\
\end{pmatrix}
\]
Consider groups $a$ and $b$ which have intra-community links and directed links from $a$ to $b$ only (so $S_{ba} = 0$) and are otherwise disconnected from the rest of the network. 

Using the inequalities above we can show $Q_{aa} > 0$, $Q_{bb} > 0$ and $Q_{ba} < 0$ always. However $Q_{ab} > 0$ only if
\begin{equation}
    S_{ab} S_{**} > S_{aa}S_{bb} 
\end{equation}
otherwise $Q_{ab} < 0$ and a simple two community structure (where $B_{ab}=-1$) is preferred. This means a community hierarchy can be detected if
\begin{itemize}
    \item The number of edges between communities, $S_{ab}$, is large.
    \item The number of edges in each community, $S_{aa}, S_{bb}$ are small.
    \item The community-hierarchy is part of a much larger network, $S_{**} \gg 0$.
\end{itemize}

\subsection{Core-Periphery Example}

To see what these limits look like in practice, consider an undirected network with three groups, a core $c$, a periphery $p$ and a disconnected community $m$. Let there be an equal number, $N$, of nodes in each group. Construct a network using the Stochastic Block Model such that every possible edge within and between groups is added with the corresponding probability from the matrix
\[
\omega = \begin{pmatrix}
    p_c & p_p & 0 \\
    p_p & 0 & 0 \\
    0 & 0 & p_m
\end{pmatrix}.
\]
These networks will be evaluated against two block patterns
\[
Q_{CP} \equiv Q \left( \begin{pmatrix}
    1 & 1 & -1 \\
    1 & -1 & -1 \\
    -1 & -1 & 1 
\end{pmatrix}\right)
\]
and
\[
Q_{Bipartite} \equiv Q \left( \begin{pmatrix}
    -1 & 1 & -1 \\
    1 & -1 & -1 \\
    -1 & -1 & 1 
\end{pmatrix}\right)
\]
where the nodes are labelled in the obvious way. If the CP structure is detectable then $Q_{CP} > Q_{Bipartite}$.

\begin{figure*}
    \centering
    \includegraphics[width=\textwidth]{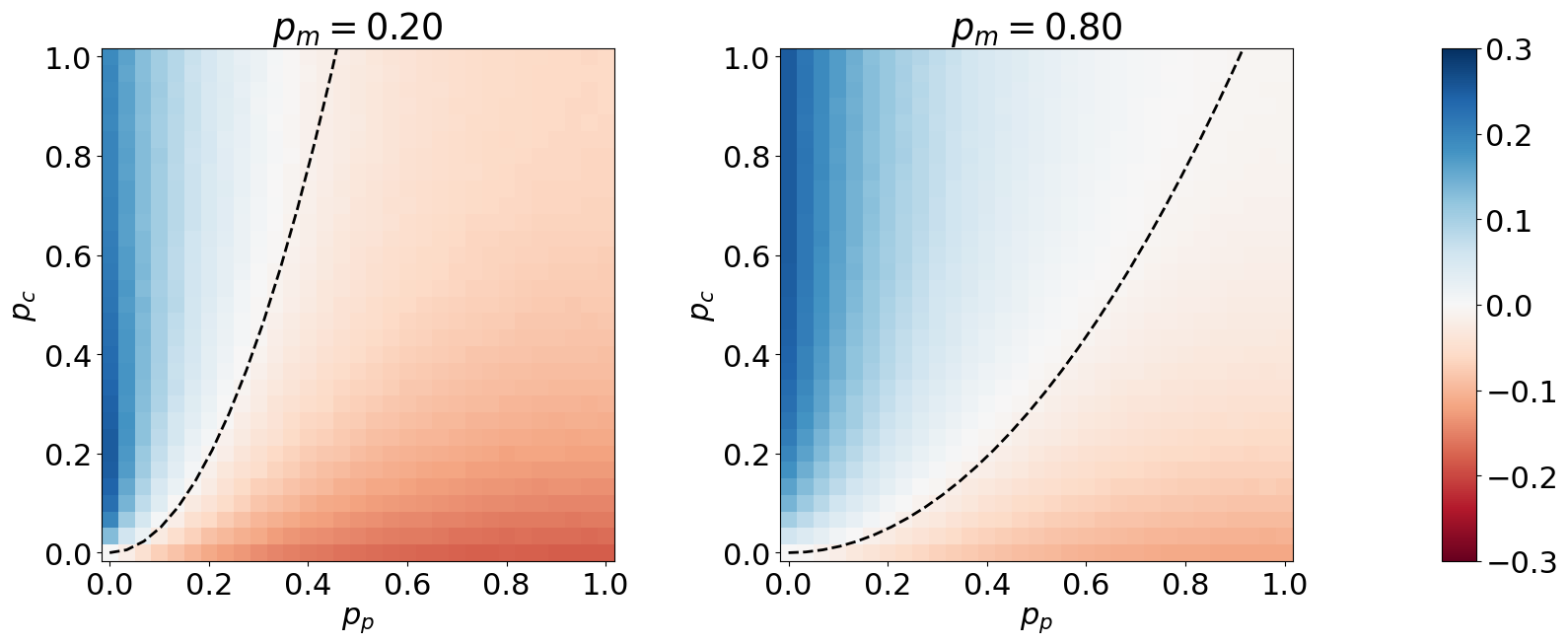}
    \caption{Color is $Q_{CP} - Q_{Bipartite}$ for fixed $p_m$ and variable $p_c, p_p$. Blue means CP can be detected under the configuration model, red means a bipartite structure is always preferred. The dashed line shows where $Q_{CP} = Q_{Bipartite}$.}
    \label{fig:cpexample}
\end{figure*}
Figure \ref{fig:cpexample} shows $Q_{CP} - Q_{Bipartite}$ for fixed $p_m$ as $p_c$ and $p_p$ vary. $N=30$ and each pair of values is averaged over 20 random networks. $p_m$ controls the size of the rest of the network and larger values give a larger region where CP structure can be detected. The dashed line shows where $Q_{CP} = Q_{Bipartite}$, corresponding to the curve $p_c = \frac{p_p^2}{p_m}$. Pairs such as  $(p_p,p_c) = (0.4,0.5)$ are notable as values of $p_c$ and $p_p$ where CP structure is detectable when the rest of the network is large ($p_m = 0.8$) and undetectable when it is small ($p_m = 0.2$).

\section{Nestedness}\label{sec:nestedness}

Nestedness is a network property often considered in the analysis of ecological networks \cite{nielsen2007ecological} but with many other applications \cite{konig2014nestedness}. A nested network has a hierarchy of degrees such that the neighbors of a node low degree node are a subset of the neighbours of any higher degree node \cite{mariani2019nestedness}. A perfectly nested unipartite network admits an ordering of the nodes that brings the adjacency matrix into a triangular form. For example, the matrix
\[
A = \begin{pmatrix}
\mathbf{0} & \mathbf{1} &  \mathbf{1} &  \mathbf{1} & 1 & 1 & 1 & 1\\
 \mathbf{1} & \mathbf{0} &  \mathbf{1} &  \mathbf{1} & 1 & 1 & 1 & 0\\
 \mathbf{1} &  \mathbf{1} & \mathbf{0} &  \mathbf{1} & 1 & 1 & 0 & 0 \\
 \mathbf{1} &  \mathbf{1} &  \mathbf{1} & \mathbf{0} & 0 & 0 & 0 & 0 \\
1 & 1 & 1 & 0 & 0 & 0 & 0 & 0 \\
1 & 1 & 1 & 0 & 0 & 0 & 0 & 0 \\
1 & 1 & 0 & 0 & 0 & 0 & 0 & 0 \\
1 & 0 & 0 & 0 & 0 & 0 & 0 & 0 
\end{pmatrix}
\]
is perfectly nested. As noted in \cite{lee2016network}, this concept is very close to the idea of a core-periphery structure. Indeed in the ecological context, \cite{nielsen2007ecological} states, `To be nested, a
network must consist of a core group of generalists all interacting with each other, and
with extreme specialists interacting only with generalist species', very similar to the definition of a core-periphery network.

By partitioning the nodes into two blocks, $c, p$, where nodes in $c$ are highly connected (the bold upper left block) and rest of the nodes are in $p$, a nested network has the same block structure as a $2\times2$ core-periphery network. In practice we would not require perfectly nested structures, only that there is an excess or deficit of links compared the the expectation under the null model. The KM rules imply that, under the configuration model, such nested matrices are not detectable. Adding other structure to the network to avoid the KM rule leads to the CP constraint, Equation \ref{eqn:cpconditions}.

Nestedness is often studied in bipartite networks. For example, interactions between animals and plants, where no animal-animal or plant-plant interactions are considered. In this case the adjacency matrix has the form
\[
A = \begin{pmatrix}
    0 & \tilde{A} \\
    \tilde{A}^T & 0
\end{pmatrix}
\]
Where $\tilde{A}$ is the $N \times M$ bi-adjacency matrix. In this context, a nested structure is one where $\tilde{A}$ has the distinct nested triangular form for some ordering of the nodes. We can split the bi-adjacency matrix into a core and periphery and evaluate the partition using the $4 \times 4$ block structure
\[
B = \begin{pmatrix}
    -1 &  -1 & 1 & 1 \\
    -1 &  -1 & 1 & -1 \\
    1 & 1 & -1 & -1 \\
    1 & -1 & -1 & -1 
\end{pmatrix}
\]
$Q(B)$ will be large and positive if the matrix is bipartite and if each part has a core that interacts with the other core and periphery and a periphery that only interacts with the other core. As this is a $4\times4$ block with a mixture of $\pm1$ in each row, the KM rules are satisfied. Denote the left part by $a$ and the right by $b$. $a$ and $b$ are then each split into core $c$ and periphery $p$ giving 4 groups: $ca, pa, cb, pb$. For this kind of structure to be detected requires that $Q_{ca,cb}, Q_{ca,pb}, Q_{cb,pa} > 0$ and the rest are less than $0$.

Assuming the best case of perfect bipartite structure and perfect CP structure between parts gives $Q_{pa,pb} < 0$, $Q_{ca,pb} < 0$ and $Q_{cb,pa} < 0$. The term $Q_{ca,cb}$ is the interesting one. Using $* = \{pa, pb\}$ gives $S_{**} = 0$, $S_{ca,*} = S_{ca, pb}$ and $S_{cb,*} = S_{cb,pa}$. Substituting into Equation \ref{eqn:fullQconditions} gives
\begin{align}
Q_{ca, cb} > 0 &\equiv S_{ca,cb}(S_{ca,cb} + S_{ca,pb} 
+ S_{cb,pa}) > 
S_{ca,pb} S_{cb,pa}
\end{align}
This condition is somewhat more complicated that the CP and CH conditions, but broadly requires
\begin{itemize}
    \item The number of interactions between the cores of $a$ and $b$, $S_{ca,cb}$, is large.
    \item The number of interactions between cores and peripheries, $S_{ca,pb},S_{cb,pa}$ is small.
\end{itemize}

These results clarify the somewhat elusive nature of nestedness as highlighted in e.g. \cite{joppa2010nestedness,jonhson2013factors, payrato2019breaking, bruno2020ambiguity}. The configuration model is quite constraining. While an observed network may be sorted so that the adjacency matrix shows nested structure, a random network with the same degree sequence could also be sorted to show similar structure. Under a less restrictive null model, like the ER model, the KM rules do not hold and nestedness is detectable, as found in e.g. \cite{joppa2010nestedness}. However, weakening the null  risks confusing degree correlations for other structures \cite{jonhson2013factors}, in particular dissassortative/bipartite ones. Work like \cite{sole2018revealing}, which situates nested structures inside larger networks is more promising in terms evading the KM rules and detecting nestedness under the configuration model.

\begin{figure*}
    \centering
    \includegraphics[width=\textwidth]{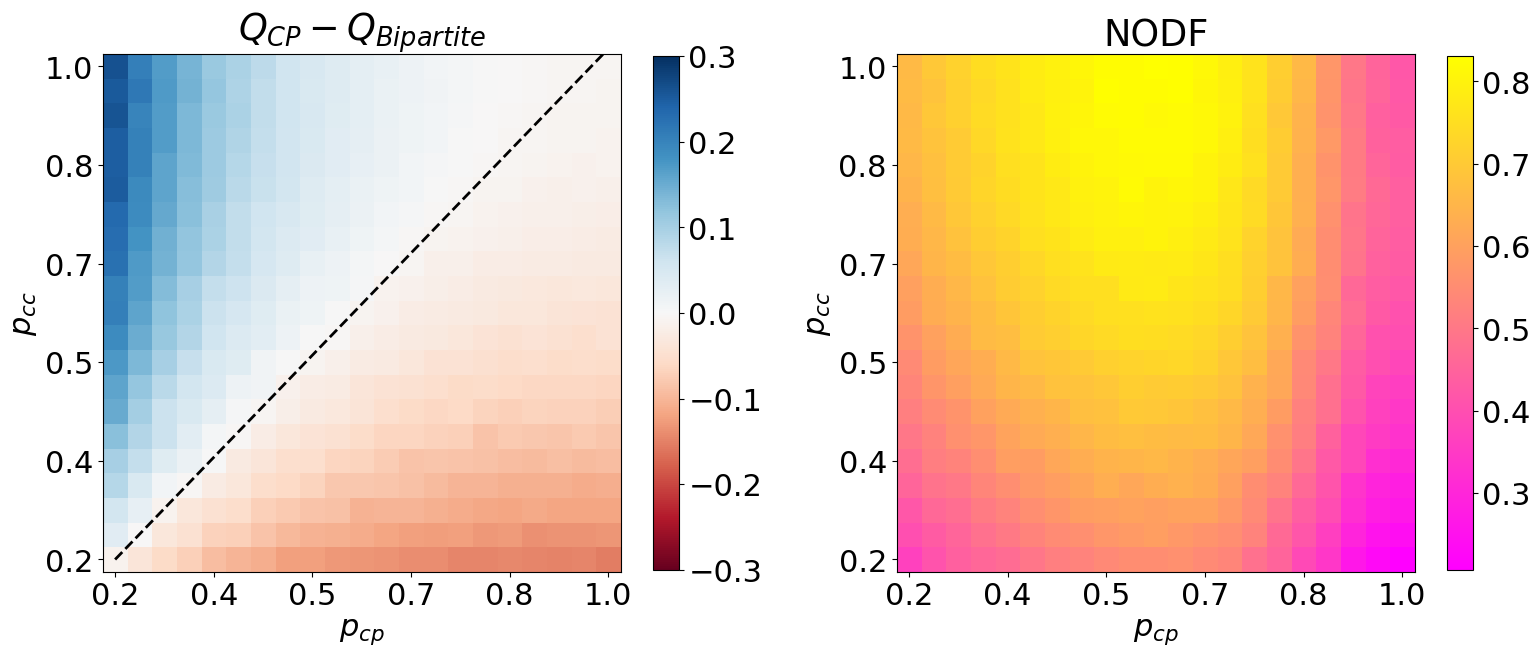}
    \caption{Left: $Q_{CP} - Q_{Bipartite}$ networks constructed as described in the text. Blue means CP can be detected under the configuration model, red means a bipartite structure is always preferred. The dotted line shows where $Q_{CP} = Q_{Bipartite}$. Right: NODF metric for the same networks. High values (near 1) imply a strongly nested network. The average of 20 random networks used for every $p_{cp}, p_{cc}$ pair. }
    \label{fig:nestexample}
\end{figure*}
Figure \ref{fig:nestexample} shows what happens for networks constructed using the SBM where the edge probabilities are given by
\[
\begin{pmatrix}
0 & 0 & p_{cc} & p_{cp} \\
0 & 0 & p_{cp} & 0 \\
p_{cc} & p_{cp} & 0 & 0 \\
p_{cp} & 0 & 0 & 0 \\
\end{pmatrix}
\]
placing $N_c = 10$ nodes in the core and $N_p = 25$ in the periphery. We evaluate block modularity against the patterns
\[
Q_{CP} \equiv Q \left( \begin{pmatrix}
    -1 & -1 & 1 & 1 \\
    -1 & -1 & 1 & -1 \\
    1 & 1 & -1 & -1 \\
    1 & -1 &-1 & -1 \\
\end{pmatrix}\right)
\]
and
\[
Q_{Bipartite} \equiv Q \left( \begin{pmatrix}
    -1 & -1 & -1 & 1 \\
    -1 & -1 & 1 & -1 \\
    -1 & 1 & -1 & -1 \\
    1 & -1 &-1 & -1 \\
\end{pmatrix}\right)
\]
and evaluate nestedness using the popular NODF metric \cite{almeida2008consistent}. The key point is that there are cases where a bipartite structure is preferred to a CP one but the NODF value is still quite high, for example, the point $(p_{cp}, p_{cc}) = (0.7,0.5)$. This would suggest that even though the network is apparently strongly nested, this is not surprising under the configuration null.

\section{Resolution Limit}\label{sec:resolution}

The well-known resolution limit of modularity \cite{fortunato2007resolution} can be derived via a similar argument. Consider two groups $a,b$ which are part of a larger network. It is better, in terms of increasing $Q(B)$, to merge $a$ and $b$ if $Q_{ab} > 0$. Under the configuration model
\begin{equation*}
Q_{ab} > 0 \equiv S_{ab} > \frac{(S_{aa} + S_{ab} + S_{a*})(S_{bb} + S_{ab} + S_{b*})}{2E}
\end{equation*}
where the $*$ index denotes all the nodes in the network other than the nodes in $a,b$. If there are no links between $a$ and $b$, then $S_{ab} = 0$ and, provided there are some internal links in $a$ or $b$, we can never have $Q_{ab} > 0$ and so the communities should always be kept separate.

The resolution limit in \cite{fortunato2007resolution} arises by considering a network with equal numbers of edges in $a$ and $b$, one link between communities $a$ and $b$, and one link connecting each of $a$ and $b$ to the rest of the network giving $S_{ab} = S_{a*} = S_{b*} = 1$ and $S_{aa} = S_{bb} = 2l$. Substituting into the above gives $Q_{ab} < 0$ if
\[
l < \sqrt{E} - 1 
\]
which is the result found in \cite{fortunato2007resolution} and derived in a similar way to the above in \cite{good2010performance}. As pointed out by \cite{good2010performance}, in the configuration model, when the network is large enough, the expectation for a single edge can be much less than 1. The resolution limit is therefore doing what it is designed to do, identifying that a single edge between communities is unusual under the null model. Equation (9) gives the most general form of the resolution limit. Disconnecting $a,b$ from the rest of the network, which is the best case, two communities will be kept separate if
\begin{equation}\label{eqn:reslimit}
S_{ab}S_{**} < S_{aa}S_{bb} - S_{ab}^2
\end{equation}

The threshold value of $S_{ab}$ depends not only what is happening in the groups $a$ and $b$, but also on what is happening in the rest of the network $S_{**}$. Seen in this light, all of the results presented above are a type of resolution limit, but for core-periphery, community-hierarchy and nested structures.  The resolution limit suggests that, for small communities in large networks, even single edge between `obvious' communities can be unexpected. The results of the previous section similarly imply that `obvious' CP and CH structures can be an expected as a consequence of the degree sequence.

\section{Relation of Block Modularity to the Stochastic Block Model}\label{sec:SBM}
The resolution limit is often partially evaded using the scaled configuration model \cite{reichardt2006statistical}
\[
P_{ij} = \gamma \frac{k_i k_j}{2E}
\]
 as a null. With this null model \cite{newman2016equivalence,young2018universality} shows that maximising $Q_{Newman}$ is equivalent to the method of maximum likelihood applied to a particular class of degree corrected Stochastic Block Models (dc-SBMs) \cite{karrer2011stochastic} called the planted partition model \emph{if $\gamma$ is chosen correctly}. It turns out that block modularity with the scaled configuration model is similarly related to the dc-SBM. In fact, with a diffrent choice of null, block modularity can be made to be identical to the dc-SBM model likelihood in general.

 In the dc-SBM, the edge probability is
 \[
 P_{ij} = \frac{k_i k_j}{2E} \omega_{ c(i), c(j) }
 \]
 where $\omega$ is a $K \times K$ matrix controlling the expected number of edges in and between each group. In \cite{newman2016equivalence} it is shown that maximising the log-likelihood for the undirected dc-SBM corresponds to maximisation of the function
 \begin{equation}\label{eqn:loglikelihood}
     \log P(G|\omega,c) = \frac{1}{2} \sum_{ij} \left( A_{ij} \log(  \omega_{ c(i), c(j) } ) -\frac{k_i k_j}{2E}  \omega_{ c(i), c(j) } \right) 
 \end{equation}
 Where $ P(G|\omega,c)$ is the probability of the observed network $G$ given $\omega$ and the partition $c$. The planted partition model is one where
 \[
 \omega_{ab} = (\omega_{in} - \omega_{out}) \delta_{ab} + \omega_{out}
 \]
corresponding to a matrix with $\omega_{in}$ on the diagonal and $\omega_{out}$ elsewhere. Substituting this into Equation \ref{eqn:loglikelihood}, \cite{newman2016equivalence} shows that maximising $Q_{Newman}$ is equivalent to maximising the likelihood, Equation \ref{eqn:loglikelihood}, provided the scaled configuration model is used with a particular value of $\gamma$. We can generalise this to block modularity. Define
 \begin{equation}\label{eqn:omegasbm}
     \omega_{ab} =  (\omega_{in} - \omega_{out})B_{ab} + \omega_{out} 
 \end{equation}
  using values for $B$ in $\{1,0\}$ in this case. The value $\omega_{in}$ is associated with the excesses and $\omega_{out}$ with the deficits. Substituting into Equation \ref{eqn:loglikelihood} rearranging and dropping some additive constants which don't affect the optimisation gives
 \begin{align}\label{eqn:equivblock}
\log P(G|\omega,c) = \frac{1}{2} \log \frac{ \omega_{in} }{ \omega_{out}} \sum_{ij} &B_{c(i), c(j) } 
\bigg( A_{ij} - \\ \nonumber
&\frac{k_i k_j}{2E} \frac{ \omega_{in} - \omega_{out} }{ \log \omega_{in}  - \log \omega_{out} } \bigg)     
 \end{align}
 Which is equivalent to block modularity with the scaled configuration model, up to the scaling factor $\log \frac{ \omega_{in} }{ \omega_{out}}$, \emph{provided}
 \[
 \gamma = \frac{ \omega_{in} - \omega_{out} }{ \log \omega_{in}  - \log \omega_{out} }.
 \]
The standard configuration model has $\omega_{in} = \omega_{out} = 1$ which, due to the overall factor $\log \frac{ \omega_{in} }{ \omega_{out}}$, causes Equation \ref{eqn:equivblock} to vanish. The likelihood, Equation \ref{eqn:loglikelihood}, is derived by computing the probability of the network $G$ given $\omega$ and $g$. With $\omega_{ab} = 1$ knowing the partition gives no extra information about the probability of an edge existing and $\log P(G|\omega,c)$
is constant.

Having one free parameter, $\gamma$, gives the model enough flexibility to be identified with the planted partition model. An even more flexible null allows maximising block modularity to be equivalent to likelihood maximisation of the dc-SBM.  The log-likelihood, Equation \ref{eqn:loglikelihood}, can be written as a sum over blocks 
\begin{equation*}
     \log P(G|\omega,c) = \frac{1}{2} \sum_{ab} \left( S_{ab} \log(  \omega_{ ab } ) -\frac{T_a T_b}{2E}  \omega_{ ab } \right) 
 \end{equation*}
Now, define a new null model, which we will call the block scaled configuration model, via
\[
 P_{ij} = \frac{k_i k_j}{2E} \gamma_{ c(i), c(j) }
\]
 Substituting into block modularity gives
 \begin{equation}\label{eqn:genmod}
    Q(B) = \frac{1}{2E} \sum_{ab} B_{ab} \left( S_{ab} -\frac{T_a T_b}{2E}  \gamma_{ ab } \right) 
 \end{equation}
If  $$B_{ab} = \log \omega_{ab}$$ and 
$$
\gamma_{ab} = \frac{\omega_{ab}}{\log \omega_{ab}}
$$
then optimising the likelihood and optimising the modularity are equivalent. Here $\omega_{ab}$ is the edge probability relative to the configuration model. For $\omega_{ab} < 1$, $B_{ab} < 0$ and $\gamma_{ab} < 0$ meaning $Q_{ab} < 0$, so this term always reduces the modularity/likelihood and optimising the partition will involve trying to place fewer edges between $a$ and $b$. For $\omega_{ab} > 1$ we have  $B_{ab} > 0$ and $\gamma_{ab} > 0$, so the modularity/likelihood can be increased by having more edges between $a$ and $b$ than the null model. This is similar to how we counted excesses and deficits in the block modularity with the configuration null, except the excesses and deficits are now weighted. 

 Despite the possibility of making modularity equivalent to the maximisation of a dc-SBM likelihood, it is probably better to think of them as conceptually different approaches. In the `inferential' approach \cite{peixoto2023descriptive}, the parameters of the dc-SBM which could have generated the observed data are determined by maximising the likelihood. In the `descriptive' approach, modularity maximisation, the observed network is compared to the expectation under some null hypothesis and unexpected partitions are highlighted. We compare these approaches on CP structure in the next section.

\subsection{CP structure in the dc-SBM}

In the modularity framework there is no requirement that $B_{ab}$ and $\gamma_{ab}$ be related. If they are assumed to be related via $\omega_{ab}$ according to Equation \ref{eqn:genmod}, the values of $\omega_{ab}$ are not known and must be estimated. We could proceed as in \cite{newman2016equivalence} by choosing some starting values of $B$ and $\gamma$; finding the partition that optimises the likelihood/modularity; estimate the dc-SBM parameters using
\begin{equation}\label{eqn:estimateomega}
\omega_{ab} = 2E \frac{ S_{ab} }{ T_a T_b }
\end{equation}
update $B$ and $\gamma$ and iterate the process until convergence. Preliminary experiments suggest this approach is quite sensitive to the starting values of $\omega$ e.g. if we start with nodes in assortative communities, it is very hard to transition to having the nodes in disassortative groups, even if this would be optimal.

An approach that is sufficient for the small networks we will study here is a based on first fixing $K$, the number of groups. Estimating the correct number of groups is a difficult problem which we do not address, see e.g. \cite{newman2016estimating}. Starting with random values of $\omega_{ab} = 1 \pm 0.1 \text{ unif}(-1,1)$ and random labels $c(i) \in \{0,\ldots,K-1\}$, Equation \ref{eqn:genmod} is maximised  using a greedy label swapping heuristic. Visiting the nodes in random order, we compute the change in $Q$ from swapping the label of $i$ to any of the other allowed labels and choosing the swap which leads to the largest increase in $Q$. When computing the change caused by a label swap we consider the change in, $S_{ab}$ and $T_a$, as well as $\omega_{ab}$ using Equation \ref{eqn:estimateomega}, and consequently the change in $B_{ab}$ and $\gamma_{ab}$.

\begin{figure*}
    \centering
    \includegraphics[width=\textwidth]{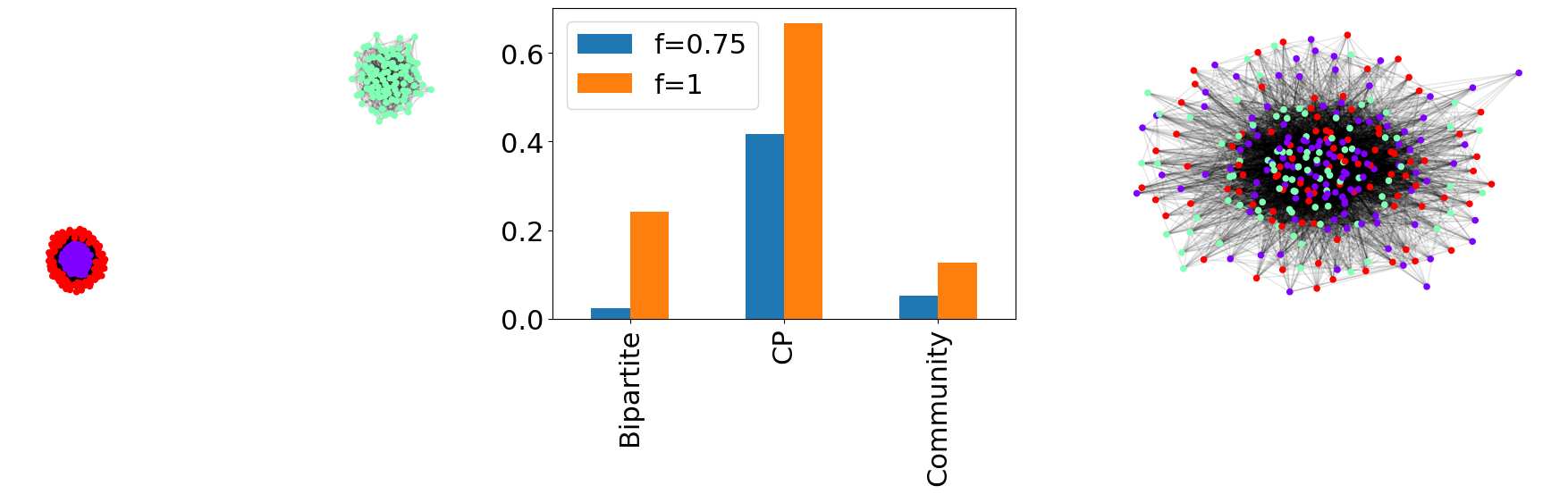}
    \caption{Left: SBM network, partition inferred by maximising the likelihood, Equation \ref{eqn:genmod}. Centre: Proportion of three structures found in 1000 samples from the configuration model of the left hand network. Right: Label assignment inferred for a configuration model network with CP structure. }
    \label{fig:configexample}
\end{figure*}

Figure \ref{fig:configexample} uses the regular SBM to generate a network with parameters
\[
\omega = \begin{pmatrix}
0.6 & 0.5 & 0\\
0.5 & 0 & 0\\
0.0 & 0 & 0.2
\end{pmatrix}
\]
The left panel shows that the optimisation of Equation \ref{eqn:genmod} recovers the expected partition into core, periphery and independent community. However note from Figure \ref{fig:cpexample} that this combination of parameters is one which is `forbidden' under the configuration model, Equation \ref{eqn:cpconditions}. Indeed, for this specific network and this partition we have $Q_{CP} = 0.54$ and $Q_{Bipartite} = 0.62$, where these are computed as in Section \ref{sec:constraint}.

This conflict arises because the two approaches are asking and answering different questions. The effectiveness of inference using the dc-SBM is generally discussed in terms of efficiently compressing the network \cite{rosvall2009map,peixoto2013parsimonious}. In our language this approach not only changes the label assignments but also the null model. The result of optimising both modularity and null is the partition and parameters that could be (and this case were) used to generate the network. In contrast, using modularity we suppose that our network is drawn at random from a fixed null model. Using the configuration model of the observed network, this means picking a random network with the same degree sequence. We then ask about the chance of finding the same structures in a random network as seen in the observed one. 

The middle panel of Figure \ref{fig:configexample} shows what happens when we do this explicitly. For the network on the left, 1000 random networks were created with the same degree sequence. The most likely $\omega$ that could have generated these networks in the dc-SBM is inferred and checked for different $2\times2$ block patterns using the following rules:
\begin{itemize}
    \item First check if: $\omega_{aa} + 2\omega_{ab} + \omega_{bb} > 2E/2$ i.e. the $2\times2$ block contains at least half of the network's edges. 
    \item For every distinct pair of communities the network has the corresponding structure if one or more of the conditions below is satisfied:
    \begin{itemize}
    \item Community: $\omega_{ab} < f\min( \omega_{aa}, \omega_{bb})$
    \item Bipartite: $f\omega_{ab} > \max( \omega_{aa}, \omega_{bb})$
    \item CP: $\min( \omega_{aa}, \omega_{bb}) < f\omega_{ab}$ and $\min( \omega_{aa}, \omega_{bb}) < f\max( \omega_{aa}, \omega_{bb})$
    \end{itemize}
\end{itemize}
$f \leq 1$ is a parameter where smaller values make the conditions more stringent. The key point is that CP structure is often found in samples from the configuration model of this network while bipartite structure is less common. This is also what modularity and the fact that $Q_{CP} < Q_{Bipartite}$ implies.

\section{Conclusions}

This analysis shows that the configuration model is quite constraining in terms of the meso-scale structures that it allows. There are only two possible $2 \times 2$ patterns, and there are strong constraints on the occurrence of CP and CH pairs as part of larger meso-scale structures. Nested networks are also quite restricted. Using a weaker null model, like the ER null, relaxes these constraints but risks confusing structure with effects better attributed to the network's degree sequence. An alternative approach, as discussed in Section \ref{sec:SBM}, is based on supposing that some flexible model, in this case the dc-SBM, generated the observed network and finding the most likely model parameters. This inferential approach has been strongly advocated \cite{peixoto2014hierarchical,guimera2020one,peixoto2023descriptive} as superseding the `descriptive' approach of modularity maximisation, paralleling vigorous debates in statistics over inferential Bayesian statistics versus classical methods \cite{efron1986isn,gelman2008objections}. 

The failure of modularity to detect the intuitive partition, the resolution limit \cite{fortunato2007resolution}, has long been discussed and is generally seen as something to be evaded either by modifying the null \cite{reichardt2006statistical} or altering the definition of modularity \cite{chen2015new,miyauchi2016z}. As shown in this work, the resolution limit is a specific example of a general type of constraint which arises when analysing network structure against a null model. Inequalities like Equations \ref{eqn:fullQconditions} show that the configuration model is quite stringent and that intuitive structures, like communities, core-periphery systems and nestedness can sometimes be explained by the degree sequence alone - any random network with that degree sequence is likely to have the `intuitive' structure, so finding it is expected and the highest modularity partition will identify something else as the most unexpected structure. 

The results of Figure \ref{fig:configexample} are striking. Maximising the dc-SBM likelihood finds the planted CP structure which modularity rejects. As discussed in Section \ref{sec:SBM}, these approaches are answering two different questions. The inference is telling us how our network could have been constructed by picking groups and randomly connecting them according to some probability matrix. Modularity is telling us how likely we are to see a given meso-structure in a random network with this degree sequence. They can and do give different answers and the most appropriate method of analysis surely depends on the questions of interest to the researcher.

There are numerous practical problems in community detection using modularity \cite{peixoto2023descriptive}, methods that address them \cite{miyauchi2016z,yanchenko2024generalized} and papers comparing the performance of different methods in practice \cite{ghasemian2019evaluating}. This work does not aim to address this. Rather the aim is to incorporate structures other than assortative communities into discussions of modularity and the resolution limit as well as to clarify the connection and the difference between modularity maximisation and model inference. It has been pointed out before that formalising the intuitive notion of `community' or `structure' is difficult \cite{good2010performance} and a universal best definition is not even possible \cite{peel2017ground}. The constraints found here hopefully shed light on the elusiveness of certain network structures and help practitioners better understand and use methods of meso-scale structure detection. 

 \appendix
 \section{Other Null Models}\label{sec:appendix}
 \subsection{Erd\H{o}s-Rényi Model}
With the ER null model the global sum provides a hard constraint
\begin{align}
\sum_{ab} Q_{ab} &= \sum_{ab} S_{ab} - \sum_{ab} p N_a N_b \\ \nonumber
    &= 2E - p N^2= 0
\end{align}
where $N_a$ is the number of nodes in group $a$. This means under the ER null model we can have an excess (or deficit) between one group and all the others (including itself), but we can't have an excess or deficit between every group with every other one. This rules out the all white or all black block patterns. The row sums are
\begin{align}\label{eqn:ERrule}
\sum_{b} Q_{ab} &= \sum_{b} S_{ab} -  N_a p \sum_{b} N_b = T_a - 2E \frac{N_a}{N}
\end{align}
So if there is a group with all excesses, this is an upper limit on the contribution of that group to the modularity, and similarly for deficits, the negative of this is a lower bound. The equivalent of Equation \ref{eqn:fullQconditions} is
\begin{equation}
    Q_{ab} > 0 \equiv \frac{ S_{ab} }{2E}  >  \frac{N_a N_b}{N^2}
\end{equation}
i.e. the fraction of edges observed between or within groups $a$ and $b$ should be greater than the fraction of edges which could possibly exist.

\subsection{Scaled Configuration Model}
The row sums for the scaled configuration model are
\begin{align}
    \sum_{b} Q_{ab} &= T_a(1 - \gamma)
\end{align}
and the total sum is
\begin{align}
    \sum_{ab} Q_{ab} &= 2E(1 - \gamma)
\end{align}
The value of $\gamma$ sets the upper and lower limits on sums of all black or all white rows. For evading the resolution limit values of $\gamma > 1$ are usually proposed \cite{lancichinetti2011limits}. However for detecting CP structure, $\gamma < 1$ is preferable.  The equivalent of Equations \ref{eqn:Qconditions} and 11 is
\begin{align}\label{eqn:sconditions}
    Q_{ab} > 0 &\equiv S_{ab}S_{**} > \gamma (S_{aa}S_{bb}-S_{ab}^2) \\ \nonumber &\qquad\qquad+ (\gamma - 1)S_{ab}(S_{aa} + 2S_{ab} + S_{bb}) \\
    Q_{ab} > 0 &\equiv S_{aa}S_{**} > \gamma (S_{ab}^2-S_{aa}S_{bb})\\ \nonumber &\qquad\qquad+ (\gamma-1)S_{aa}(S_{aa} + 2S_{ab} + S_{bb} ) 
\end{align}
The scaled configuration model can be treated by substituting $\gamma \rightarrow \gamma_{ab}$ in the equations above. For CP structure, requiring $Q_{cc} > 0$ is equivalent to
\begin{align}
    Q_{cc} > 0 &\equiv S_{cc}S_{**} > \gamma S_{cp}^2 + (\gamma-1)S_{cc}(S_{cc} + 2S_{cp}) 
\end{align}
This means $\gamma < 1$ makes CP structure easier to detect under this model. Similar results to the ones found in Sections \ref{sec:nestedness} and \ref{sec:resolution} for nestedness and the resolution limit can be derived using the above equations.


\bibliography{sample}

\end{document}